%% file: paper.tex
\documentclass[11pt]{article}
\usepackage{graphicx}

\newcommand{\BABARPubYear}    {05}

\newcommand{\BABARConfNumber} {020}
\newcommand{\SLACPubNumber} {11390}

\input babarsym

\long\def\inst#1{\par\nobreak\kern 4pt\nobreak
    {\it #1}\par\vskip 10pt plus 3pt minus 3pt}

\def\Bztokspp{\ensuremath{B^0\to\KS\pi^0\pi^0}\xspace}
\def\scp{\ensuremath{S_{\KS\pi^0\pi^0}}\xspace}
\def\ccp{\ensuremath{C_{\KS\pi^0\pi^0}}\xspace}
\def\sf{\ensuremath{S}\xspace}
\def\cf{\ensuremath{C}\xspace}
\def\finalscp{\ensuremath{0.84\pm 0.71\, (\mbox{\small stat}) \pm 0.08\, 
(\mbox{\small syst})} \xspace}
\def\finalccp{\ensuremath{0.27\pm 0.52\, (\mbox{\small stat}) \pm 0.13\, 
(\mbox{\small syst})} \xspace}

\def\Btag{\ensuremath{B_{\rm tag}}\xspace}
\def\Brec{\ensuremath{B_{CP}}\xspace}
\def\Bcp{\ensuremath{B_{CP}}\xspace}

\begin{document}
{\pagestyle{empty}

\begin{flushleft}
\end{flushleft}

\begin{flushright}
\babar-CONF-\BABARPubYear/\BABARConfNumber \\
SLAC-PUB-\SLACPubNumber \\
\end{flushright}

\par\vskip 3cm

\begin{center}
\Large \bf Measurement of \boldmath{\CP} Asymmetries in 
\boldmath{$B^0\to \KS \pi^0\pi^0$} Decays
\end{center}
\bigskip

\begin{center}
\large The \babar\ Collaboration\\
\mbox{ }\\
\today
\end{center}
\bigskip \bigskip

\begin{center}
\large \bf Abstract
\end{center}
We present a preliminary measurement of the time-dependent \CP 
asymmetry for the neutral $B$-meson decay into the $CP = +1$ 
final state $\KS\pi^0\pi^0$,
with $\KS\to \pi^+\pi^-$ and $\pi^0\to\gamma\gamma$. 
We use a sample of approximately 227 million $B$-meson pairs recorded 
at the $\Upsilon(4S)$ resonance by the \babar\ detector at the \pep2\ 
$B$-Factory at SLAC. From a maximum
likelihood fit we extract the mixing-induced \CP-violation
parameter $\scp = \finalscp$ and the direct \CP-violation
parameter $\ccp = \finalccp$, where the first uncertainty
is statistical and the second systematic.
\vfill
\begin{center}
Presented at the 
International Europhysics Conference On High-Energy Physics (HEP 2005),
7/21---7/27/2005, Lisbon, Portugal
\end{center}

\vspace{1.0cm}
\begin{center}
{\em Stanford Linear Accelerator Center, Stanford University, 
Stanford, CA 94309} \\ \vspace{0.1cm}\hrule\vspace{0.1cm}
Work supported in part by Department of Energy contract DE-AC03-76SF00515.
\end{center}

\newpage
} 

\input authors_conf05013.tex

\section{INTRODUCTION}
\label{sec:Introduction}
\CP violation effects in decays of $B$ mesons that are dominated by
$\b\to\s\qbar\q$ transitions, where $\q=\u,\d,\s$, are
potentially sensitive to contributions from physics beyond the
Standard Model ~\cite{ref:newphysics}. The \B{}-factory
experiments have explored time-dependent \CP{}-violating (CPV)
asymmetries in several such decays~\cite{cc}, including $B^0\to\phi
K^0${}~\cite{Abe:2003yt,Aubert:2004ii},
$B^0\to\KS\KS\KS$~\cite{ksksks}, $B^0\to\eta^\prime
\KS${}~\cite{Abe:2003yt,Aubert:2003bq}, $B^0\to K^+ K^-
\KS${}~\cite{Abe:2003yt,Aubert:2004ta}, $\Bz\to
f_{0}(980)\KS$~\cite{Aubert:2004am} and
$B^0\to\KS\piz$~\cite{Aubert:2004xf}.  Within the Standard Model the 
asymmetry in
these decays is expected to be consistent with the asymmetry in
$\b\to\c\cbar\s$ decays, such as $\Bz \to \jpsi
\KS$, where the CPV asymmetry occurs due to a phase difference between
mixing and decay amplitudes. These comparisons must take into account
contributions of other amplitudes with different
weak-interaction phases within the Standard Model. 
A major goal of the $B$-factory experiments
is to reduce the experimental uncertainties of these measurements and
to add more decay modes in order to improve the sensitivity to
beyond-the-Standard-Model effects.

In this letter we present a preliminary measurement of the CPV asymmetry
in the decay $B^0\to\KS\piz\piz$, using data collected with the \babar{}
detector at the \pep2{} asymmetric-energy \epem{} collider. 
In the Standard Model this decay is dominated by the $b\to \s\qbar\q$ amplitude, with $q=u,d$. 
A possible contribution from a tree-level
$b\to u\bar{u}s$ amplitude is doubly Cabibbo-suppressed with respect
to the leading gluonic penguin diagram.

The $\KS\pi^0\pi^0$ final state is
a \CP -even eigenstate, regardless of any resonant
substructure~\cite{Gershon:2004tk}.
In the Standard Model we expect $\scp\simeq -\sin 2\beta$ and $\ccp \simeq 0$.
The angle $\beta$ is defined as $\beta$ = $arg(- V_{cd}
V^*_{cb}/V_{td}V^*_{tb})$ and $V_{ij}$ are the elements of the CKM
matrix~\cite{ckm}.
A significant measurement of \CP violation in this channel
alone in comparison to other penguin modes constrains 
certain extensions of the Standard Model~\cite{grossman}.

\section{THE \babar\ DETECTOR AND DATASET}
\label{sec:babar}
The data in this analysis were collected with the 
\babar\ detector~\cite{ref:babar} at the PEP-II asymmetric 
$e^+e^-$ collider ~\cite{pep}.  A sample of
$226.6\pm2.5$ million \BB\ pairs was recorded at the $\Upsilon (4S)$
resonance (center-of-mass energy $\sqrt{s}=10.58\ \gev$). 
 The \babar\ detector is described in detail
elsewhere~\cite{ref:babar}. Charged particles are detected 
and their momenta measured by the combination of a silicon 
vertex tracker (SVT), consisting of five layers
of double-sided detectors, and a 40-layer central drift chamber,
both operating in the 1.5 T magnetic field of a solenoid.
Charged-particle identification (PID) is provided by the average energy
loss in the tracking devices and by an internally reflecting ring-imaging
Cherenkov detector (DIRC) covering the central region.
Photons and electrons are detected by an electromagnetic 
calorimeter composed of 6580 CsI(Tl) crystals; the typical resolution 
for the $\piz$ signal in the $\gamma\gamma$ invariant mass spectrum is 
better than 7~\mevcc.

\section{ANALYSIS METHOD}
\label{sec:Analysis}
In the decay \Bztokspp, which has no charged tracks originating from
the \Bz decay vertex, we rely on 
the technique recently developed to reconstruct
the \Bz vertex in $B^0\to\KS\piz$ decays (described in detail below)~\cite{Aubert:2004xf}. 
From a candidate \BB\ pair we reconstruct a \Bz\  decaying into the \CP
eigenstate $\KS \pi^0\pi^0 $ ($B_{CP}$).  We also reconstruct the vertex of
the other $B$ meson ($B_{\rm tag}$) and identify its flavor.
The difference $\deltat \equiv t_{\CP} - t_{\rm tag}$
of the proper decay times is obtained from the measured distance between the $B_{\CP}$
and  $B_{\rm tag}$ decay vertices and from the boost ($\beta \gamma =0.56$) of 
the \epem system. The \deltat\ distribution is given by:
\begin{eqnarray}
\label{eq:cpt}
  {\cal P}_\pm(\Delta t) &=& 
        \frac{e^{-\left|\deltat\right|/\tau}}{4\tau} [1 \mp\Delta w \pm
                                                   \label{eq:FCPdef}\\
   &&\hspace{-1em}(1-2w)\left( \sf\sin(\deltamd\deltat) -
\cf\cos(\deltamd\deltat)\right)].\nonumber
\end{eqnarray}
The upper (lower) sign denotes a decay accompanied by a \Bz (\Bzb) tag,
$\tau$ is the mean $\Bz$ lifetime, $\deltamd$ is the mixing
frequency, and the mistag parameters $w$ and
$\Delta w$ are the average and difference, respectively, of the probabilities
that a true $\Bz$\ is incorrectly tagged as a $\Bzb$\ or vice versa.
The tagging algorithm \cite{s2b} has seven mutually exclusive tagging 
categories of
differing purities (including one for untagged events that we
retain only for yield determinations).  The analyzing power, defined as 
efficiency times $(1-2w)^2$ summed over all categories, is $( 30.5\pm 0.6)\%$,
as determined from a large sample 
of $B$-decays to fully reconstructed flavor eigenstates ($B_{\rm flav}$).

We search for \Bztokspp decays in \BB{} candidate events
selected using charged-particle multiplicity and event
topology~\cite{ref:Sin2betaPRD}.  We reconstruct $\KS\to\pip\pim$
candidates from pairs of oppositely charged tracks.  The two-track
combinations must form a vertex with a $\chi^2$ probability greater
than $0.001$ and a $\pip\pim$ invariant mass within
11.2~\mevcc of the nominal \KS\ 
mass~\cite{Eidelman:pdg2004}.    We form $\piz\to\gamma\gamma$
candidates from pairs of photon candidates in the EMC, each of which
is isolated from any charged tracks, carries a minimum energy of
30~\mev, and has the expected lateral shower shape.
Candidates for \Bztokspp are formed from $\KS\piz\piz$ combinations
and constrained to originate from the \epem{} interaction point using
a geometric fit. We require that the $\chi^2$ consistency of the 
fit, which has one degree of freedom, be greater than $0.001$.  We
extract the \KS{} decay length $L_{\KS}$ and the $\piz\to\gamma\gamma$
invariant mass from this fit and require 110 $< m_{\gamma\gamma}
<$ 160~\mevcc and $L_{\KS}$ greater than $5$ times its uncertainty.
The cosine of the angle between the direction of the decay photon 
in the center-of-mass system of the mother \piz and the \piz  flight
direction must be less than 0.92.

We extract the signal yield, \sf and \cf
from an unbinned extended maximum likelihood 
fit where we parameterize the distributions of several
kinematic and topological variables for signal and background events
in terms of probability density functions (PDFs).

For each $B$ candidate we compute two kinematic variables, 
the energy difference $\Delta E = E_B^* - \frac{1}{2}\sqrt{s}$ and
the beam-energy--substituted mass 
$\mes = \sqrt{(\frac{1}{2}s + \vec{p}_0\cdot\vec{p}_B)^2/E_0^2 - p_B^2}$
\cite{ref:babar}, where $s$ is the center-of-mass energy squared. The 
subscripts 0 and $B$ refer to the initial 
$\Upsilon(4S)$ and the $B_{CP}$ candidate, respectively, and the 
asterisk denotes the center-of-mass frame.
For signal events, \DeltaE is expected to peak at zero and 
\mes at the known $B$ mass. From a detailed simulation  we
expect  a signal resolution of about 3.6~\mevcc in \mes and 45~MeV in
$\Delta E$. Both distributions exhibit a 
low-side tail due to the response of the EMC to
photons. We remove a small dependence 
of the signal $\Delta E$ resolution on the location in the
$\KS\piz\piz$ Dalitz plot
by using $\Delta E/\sigma(\Delta E)$ instead of $\Delta E$, where
$\sigma(\Delta E)$ is the measured uncertainty in $\Delta E$. We select
candidates with $\mes > 5.20$~\gevcc and $-5 < \Delta E/\sigma(\Delta E) < 2$.
To suppress other $B$ decays we also require $-0.25 < \Delta E < 0.1$~GeV,
which does not affect the signal $\Delta E/\sigma(\Delta E)$ distribution.

The background $B$ candidates come primarily from random combinations 
of \KS and neutral pions produced in events of the type $e^+e^-\to q\bar{q}$, where  
$q = u,d,s,c$ (continuum). 
Background from  $\BB$ events may occur either in charmless decays
$\Bz \to \KS X$, or from decays where the \KS is from an intermediate
charmed particle.  
The shapes of event variable distributions are obtained from signal 
and background Monte Carlo (MC) samples and high statistics data control samples.
In \mes, the charmless $B$ background exhibits a broad enhancement
near the 
$B$-meson mass while other $B$ background distributions show no
peaking. In $\Delta E/\sigma(\Delta E)$, $B$ backgrounds in general
show no clustering.

In continuum events, particles appear mostly in two jets. This topology 
can be characterized with several variables computed in the 
$\Upsilon(4S)$ frame.
One such quantity is the angle $\theta_T$ between the thrust axis of the 
$B_{CP}$ candidate and the thrust axis formed from the other charged and 
neutral particles in the event, where the thrust axis is defined as the
axis that maximizes the sum of the magnitudes of the longitudinal momenta.
This angle is small for continuum events and uniformly distributed for
true $B\bar{B}$ events. With the requirement $|\cos\theta_T| < 0.9$
we suppress background by a factor of three while retaining 90\% of
the signal. We also use the angle $\theta_B$ between the $B_{CP}$ momentum 
and the beam axis, and the sum of the momenta $p_i$ of the other charged and
 neutral  particles in the event weighted by the Legendre polynomials 
$L_0(\theta_i)$ and $L_2(\theta_i)$ where $\theta_i$ is the angle between the 
momentum of particle $i$ and the thrust axis of the $B_{CP}$ candidate.
We combine these three variables in a neural net ($NN$) that is trained
and evaluated \cite{bfgs} on different subsets of simulated signal and 
continuum events and on data taken about 40~MeV below the nominal 
center-of-mass energy. The $NN$ has two hidden layers with 4 neurons each.  
The $NN$ output is divided into 10 consecutive intervals, chosen such that
they are uniformly populated by the signal events;
the PDF is modeled as a parametric step function~\cite{btopipi}
whose parameters are the heights of each bin. Since the parent distribution
for the $NN$ output is unknown any assumed functional form will suffer a 
systematic uncertainty due to the choice of the function.

We suppress background from other $B$ decays by excluding several invariant 
mass  intervals: $m(\KS\piz) > 4.8$~\gevcc eliminates $B^0\to\KS\piz$, 
$1.75 <  m(\KS\piz) < 1.99$~\gevcc reduces $B^0\to\bar{D}^0\piz$ 
to fewer than 10 expected candidates, $m(\piz\piz) < 0.6$~\gevcc removes 
$\eta\KS$ and $\eta^\prime \KS$, and $3.2 < m(\piz\piz) < 3.5$~\gevcc removes
$\chi_{c0}\KS$ and $\chi_{c2} \KS$ candidates.

From MC simulation we expect more than one candidate in 13\%
of the signal candidate events. Because the number of multiple \KS
candidates is negligible (less than 0.1\%), we select the candidate whose two 
reconstructed \piz masses are closest to the expected value.  The 
signal reconstruction efficiency is about 15\%. 

For each \Bztokspp candidate we examine the remaining tracks 
in the event to determine the decay vertex position and the flavor 
of \Btag. We parameterize the performance of the tagging algorithm 
in a data sample ($B_{\rm flav}$) of fully reconstructed $\Bz\to
D^{(*)-}\pip/\rho^+/a_1^+$ decays.
For the continuum background, the fraction of events tagged in 
category $k$, $\epsilon_k$, is extracted from a fit to the data. 
The \Btag vertex is reconstructed inclusively from the 
remaining charged particles in the event~\cite{ref:Sin2betaPRD}. 

To reconstruct the \Brec vertex from the single \KS trajectory we exploit 
the knowledge of the average interaction point (IP), which is determined 
every 10 minutes from the spatial distribution of vertices from 
two-track events. The uncertainty on the IP position, which follows
from the size of the interaction region, is about 150~$\mu$m  
horizontally and 4~$\mu$m vertically.
We compute \deltat and its uncertainty from a geometric fit~\cite{treefit} to 
the $\Upsilon(4S)\to\Bz\Bzb$ system that takes this IP constraint into
account. We further improve the sensitivity to \deltat by constraining the 
sum of the two $B$ decay times ($t_{\CP}+t_{\rm tag}$) to be equal to 
$2\:\tau_{\Bz}$ with an uncertainty of $\sqrt{2}\; \tau_{\Bz}$, which effectively 
constrains the two vertices to be near the $\Upsilon(4S)$ line of flight. 
This procedure provides an unbiased estimate of \deltat.
The extraction of \deltat with the IP-constrained fit has been extensively
tested on large samples of simulated \Bztokspp decays with different values
of \sf and \cf, and in data~\cite{Aubert:2004xf}.

The per-event estimate of the uncertainty on \deltat{} reflects the
strong dependence of the \deltat resolution on the $\KS$ flight
direction and on the number of SVT layers traversed by the $\KS$ decay
daughters. In about 70\% of the events both pion tracks 
are reconstructed from at least 4 SVT hits, leading to sufficient resolution
for the time-dependent measurement. The average \deltat{} resolution
in these events is about 1.0~ps. For events that fail this
criterion or for which $\sigma(\Delta t) > 2.5$~ps or
$\deltat>20$~ps, the \deltat information is not used.
However, since \cf can also be extracted from flavor tagging
information alone, these events still contribute to the measurement 
of \cf.

By exploiting regions in data that are dominated by background,
and simulated events for the signal, we have verified that with
our selection the observables are sufficiently independent that 
we can construct the likelihood from the product of one-dimensional 
PDFs, apart from the signal \mes and $\Delta E/\sigma(\Delta E)$
which are correlated away from their mean signal
positions and for which we use a two-dimensional 
PDF derived from a smoothed, simulated distribution.
We obtain the PDF for the \deltat of signal events from the
convolution of Eq.(\ref{eq:cpt}) with a resolution function ${\cal
  R}(\delta t \equiv \deltat -\deltat_{\rm true},\sigma_{\deltat})$.
The resolution function is parameterized as the sum of two Gaussians
with a width proportional to the reconstructed $\sigma_{\deltat}$, and
a third Gaussian with a fixed width of
8~ps~\cite{ref:Sin2betaPRD}. The first two Gaussian distributions have a
non-zero mean, proportional to $\sigma_{\deltat}$, to account for the
charm decays on the \Btag side.  We
have verified in simulation that the parameters of ${\cal R}(\delta t,
\sigma_{\deltat})$ for \Bztokspp events are similar to those
obtained from the $B_{\rm flav}$ sample, even though the distributions
of $\sigma_{\deltat}$ differ considerably. We therefore extract these
parameters from a fit to the $B_{\rm flav}$ sample. 
We use the same resolution function for background from other 
charmless $B$ decays.
The \deltat distributions for background from $B$ decays into 
open charm final states and continuum consist of a prompt component and 
a non-prompt component, and the resolution function has the same
functional form as used for signal events.
The parameters for the \deltat PDF of the open-charm background are 
determined from MC simulation, while for the continuum they 
are varied in the fit to data.

\section{MAXIMUM LIKELIHOOD FIT}
We subdivide the data into the tagging categories $k$, events with and 
without \deltat information (set $I$ and $II$), and those located in the inside or 
outside region of the Dalitz plot ($inside$ or $outside$). The latter accounts for the  
higher contribution and different characteristics of continuum background near the Dalitz 
plot boundary. We define the quantity $\delta = min(m_{12}^2,m_{13}^2,m_{23}^2)$,
where $m_{ij}$ is the invariant mass of the $B$ decay daughters $i$ and $j$
combined. It corresponds to the distance of an event in the Dalitz plot to the nearest 
Dalitz plot boundary in the limit of massless daughters. We split the data 
at $\delta = 3.5$~GeV$^2/$c$^4$. We maximize the logarithm of the extended likelihood
$\mathcal{L} = e^{(N_S + N_B)} \cdot \prod_{k}^{7} l_k$
with $N_S$ and $N_B$ ($= \sum_B n_B$) the total signal and background yields, 
respectively. The likelihood in each tagging category $k$ (with tagging fraction 
$\epsilon_k$) is given as:
\begin{eqnarray}
l_k & = & \prod_{j}^{N\mbox{\tiny I outside k}} \left[
N_S\, \epsilon^S_k f^S_g f^S_{out}\, P^S_{k,j} +
\sum_B n_B\, \epsilon^B_k f^B_g f^B_{out}\, P^B_{k,out,j}  \right] \times \nonumber \\
    &   & \prod_{j}^{N\mbox{\tiny I inside k}} \left[
N_S\, \epsilon^S_k f^S_g (1 - f^S_{out})\, P^S_{k,j} +
\sum_B n_B\, \epsilon^B_k f^B_g (1 - f^B_{out})\, P^B_{k,in,j}  \right] \times \nonumber \\
    &   & \prod_{j}^{N\mbox{\tiny II outside k}} \left[
N_S\, \epsilon^S_k (1 - f^S_g) f^S_{out}\, Q^S_{k,j} +
\sum_B n_B\, \epsilon^B_k (1 - f^B_g) f^B_{out}\, Q^B_{k,out,j}  \right] \times \nonumber \\
    &   & \prod_{j}^{N\mbox{\tiny II inside k}} \left[
N_S\, \epsilon^S_k (1 - f^S_g) (1 - f^S_{out})\, Q^S_{k,j} +
\sum_B n_B\, \epsilon^B_k (1 - f^B_g) (1 - f^B_{out})\, Q^B_{k,in,j}  \right]  \, .
\end{eqnarray}
The probabilities $P^S$ ($Q^S$) and $P^B$ ($Q^B$) for each measurement $j$ are the 
products of PDFs for signal ($S$) and background ($B$) classes:
$P_{k,j} = PDF(m_{ES j}, \Delta E/\sigma(\Delta E)_j)\cdot PDF(NN_j)\cdot 
PDF(\Delta t_j,\sigma(\Delta t)_j, \mbox{tag}_{k,j}, k_j)$,
where for the background  $PDF(m_{ES j}, \Delta E/\sigma(\Delta E)_j)$
= $PDF(m_{ES j})\cdot PDF(\Delta E/\sigma(\Delta E)_j)$.
The probabilities $Q$ do not depend on $\Delta t$ and $\sigma (\Delta t)$
and are used to extract \cf from the yields.
The fractions of events with \deltat information for signal and background 
are denoted by $f^S_g$ and $f^B_g$, respectively, and fractions of events 
in the outside Dalitz plot region by $f^S_{out}$ and $f^B_{out}$. 
For about 22\% of our signal $B$ candidates one or two of the \piz
decay photons associated with \Bcp originate from the \Btag . According to Monte Carlo 
simulation studies in these cross-feed events we expect to measure the same 
\sf and \cf as in the correctly reconstructed signal ($true$) since the contribution
of the \piz to the \deltat measurement is marginal. To account for differences
in the PDF distributions for the signal probabilities $P^S$ ($Q^S$) we use: 
$P = f_{cf} P_{cf} + (1 - f_{cf}) P_{true}$. The fraction of cross-feed events,
$f_{cf}$, is fixed to the value obtained from the simulation.
Parameters of signal PDFs are the same for the different Dalitz plot regions.
The PDFs for $B$ backgrounds are identical for the Dalitz inside and
outside regions.
The tagging fractions for the signal and the $B$ decay backgrounds 
are the same; continuum background has different $\epsilon_k^{B}$.
The good fractions $f^S_g$ and $f^B_g$ and the outside fractions
$f^S_{out}$ and $f^B_{out}$ for continuum are varied in the fit,
while these fractions for charm and charmless $B$ backgrounds
are determined from Monte Carlo simulations.
The fit was tested with both a parameterized simulation of a large number 
of data-sized experiments and a full detector simulation. 

\section{PHYSICS RESULTS}
\label{sec:Physics} 
\begin{figure}[!htb]
\begin{center}
\begin{tabular}{cc}
\includegraphics[height=5.6cm]{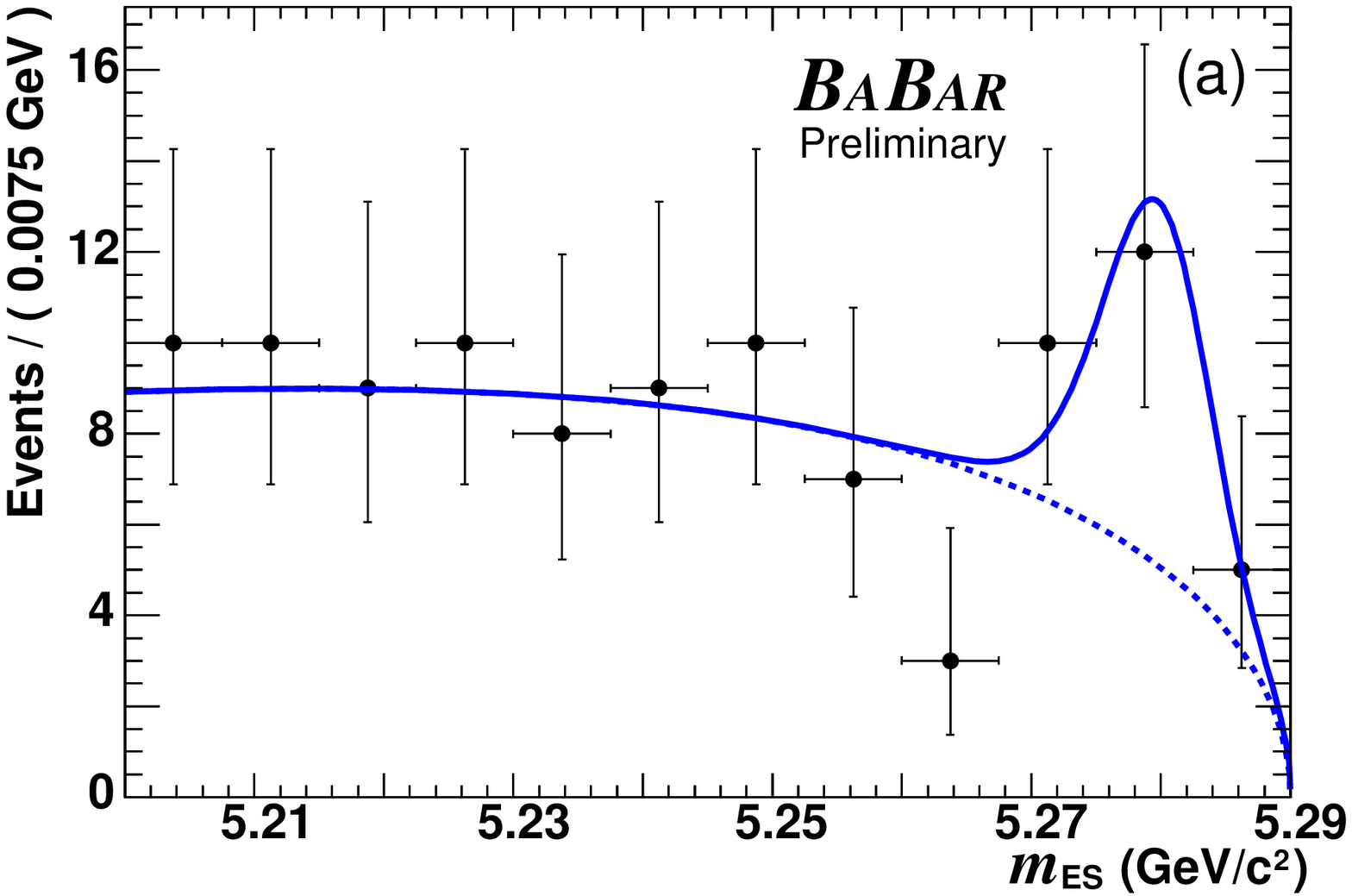} &
\includegraphics[height=5.6cm]{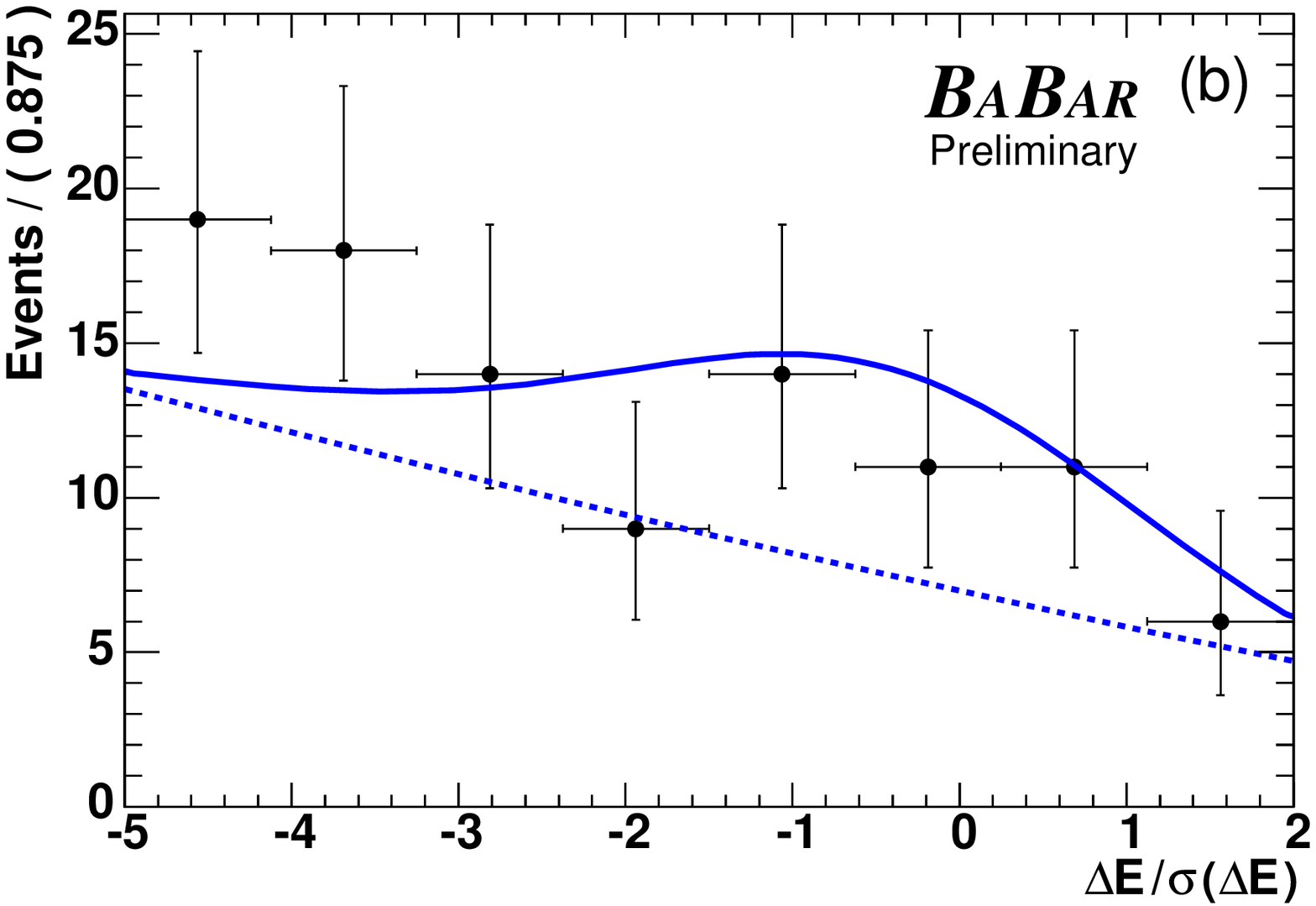} \\
\includegraphics[height=5.6cm]{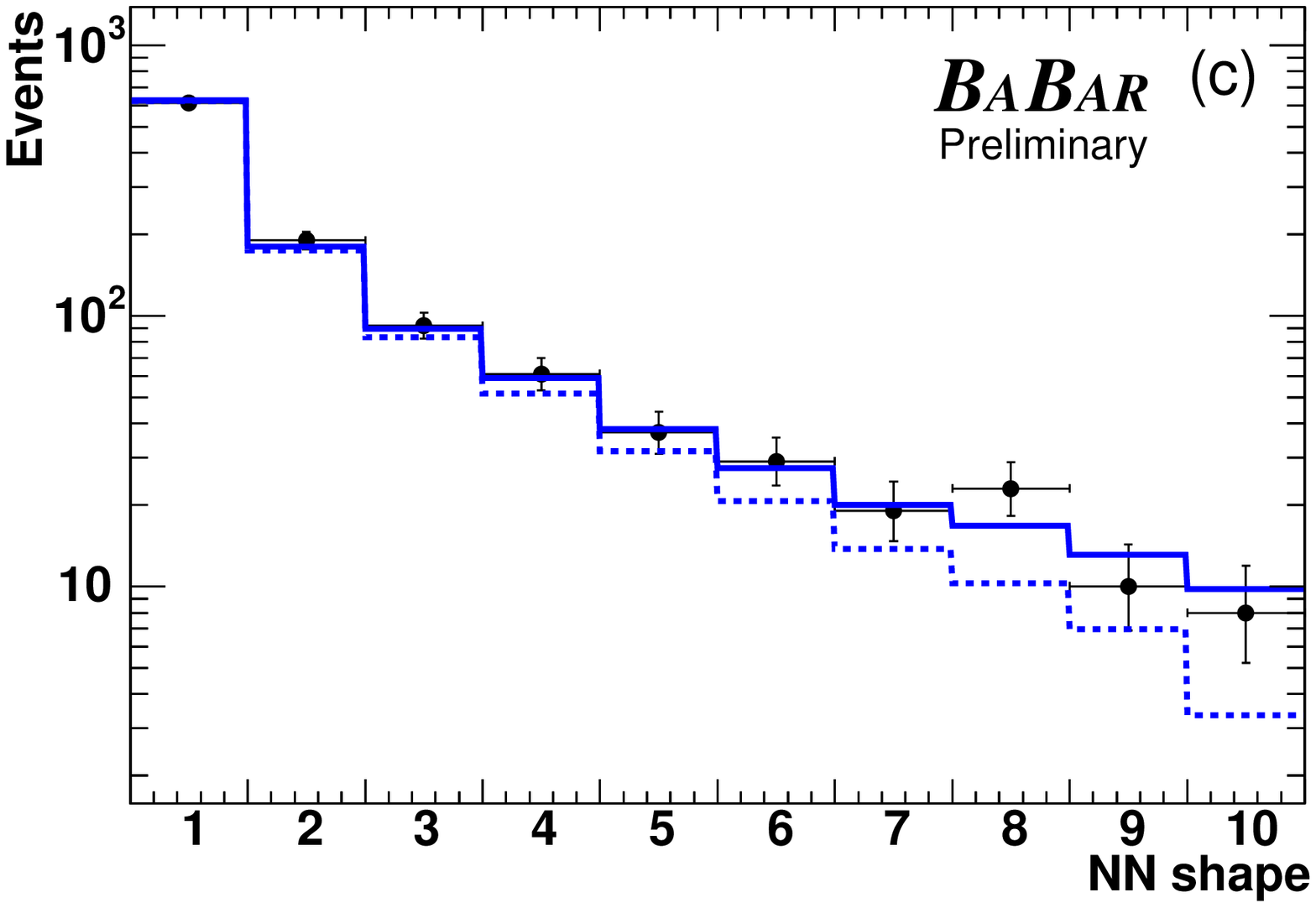} & 
\includegraphics[height=5.6cm]{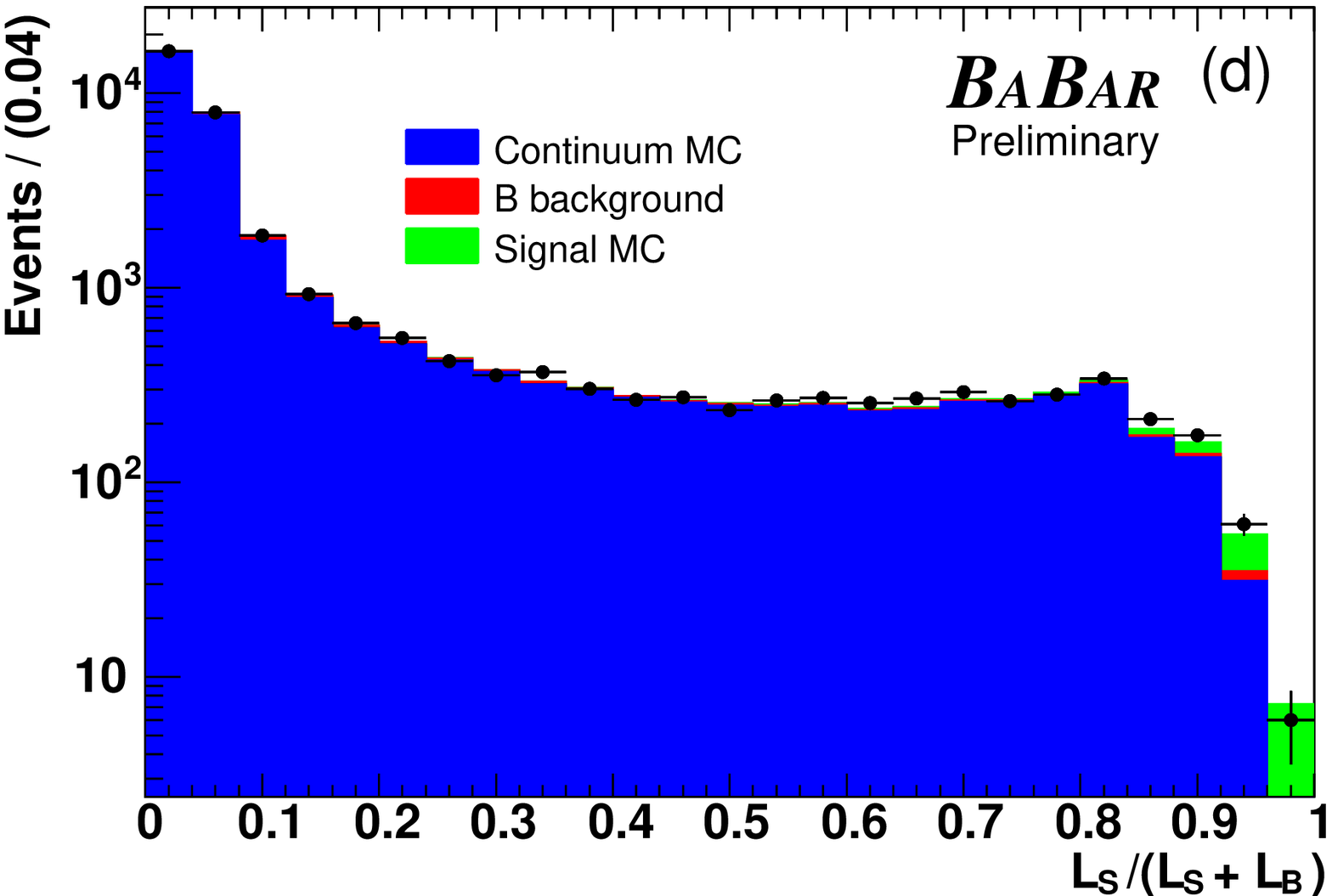} \\
\includegraphics[height=5.6cm]{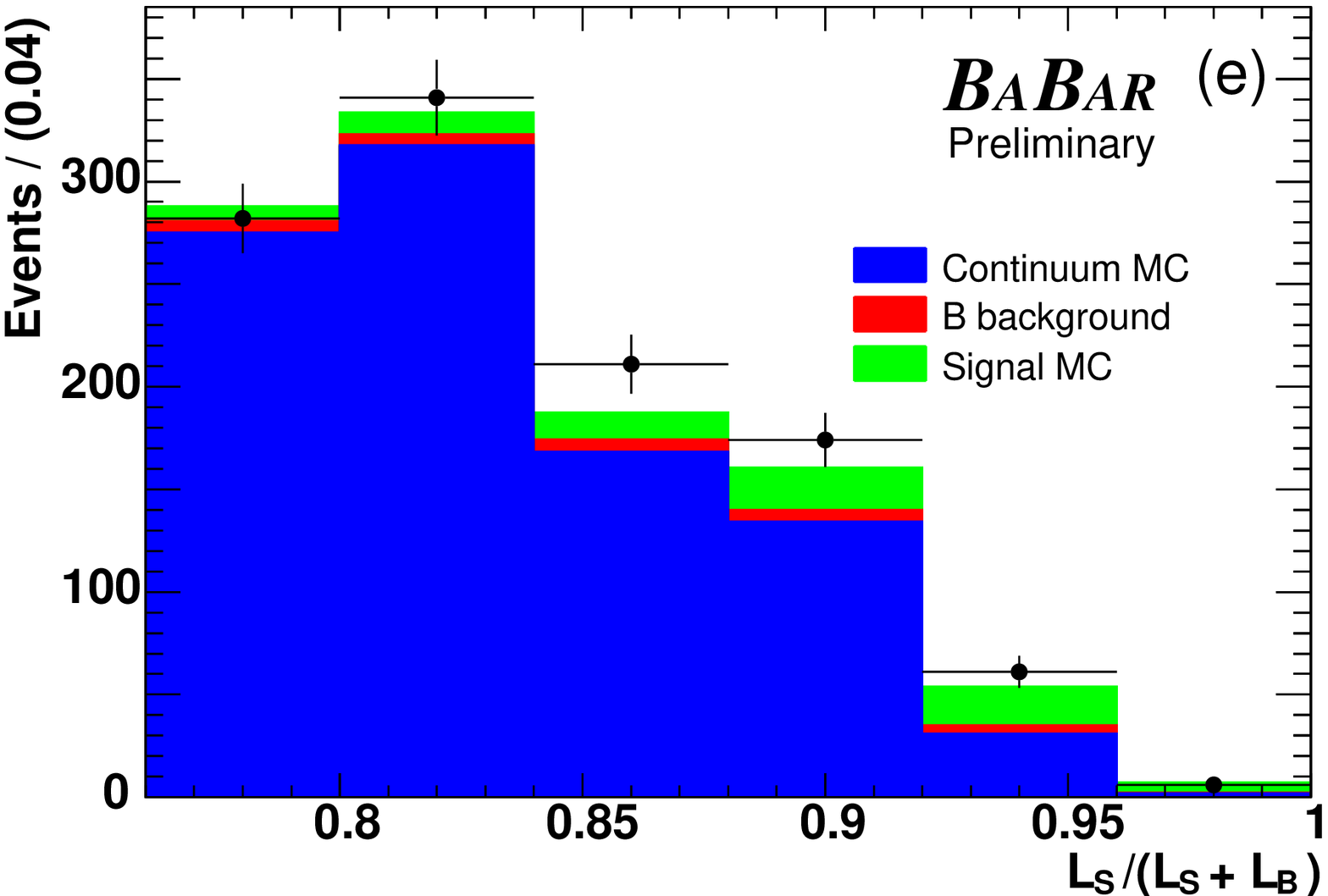} & \\
\end{tabular}
\caption{Distribution of the event variables (a) \mes, (b) $\Delta E/\sigma(\Delta E)$, 
and (c) $NN$ output in 10 bins after reconstruction and a requirement 
on the ratio of signal likelihood to the signal-plus-background likelihood, calculated 
without the plotted variable. The solid line represents the fit result
for the total event yield and the dotted line for the total background.  
Plot (d) shows the ratio of the signal likelihood to signal-plus-background
likelihood with all variables included, data (dots) with the fit 
result superimposed. Plot (e) shows the same quantity as (d) 
close to one and with a linear scale. 
\label{fig1}
}
\end{center}
\end{figure}

\begin{figure}[!htb]
\begin{center}
\includegraphics[height=14cm]{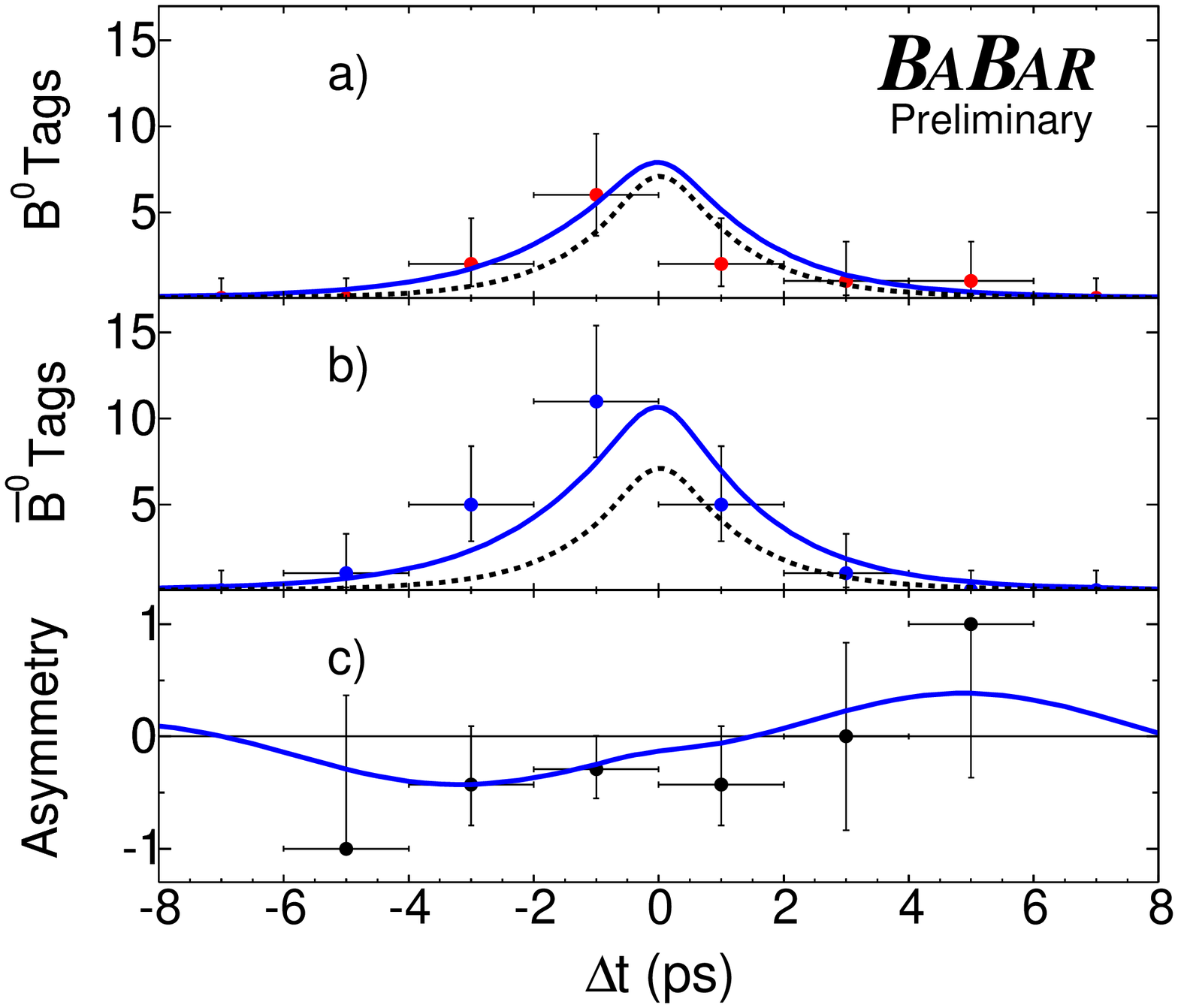} 
\caption{
Plots (a) and (b) show the \deltat distributions of $B^0$- and
$\bar{B}^0$-tagged \Bztokspp candidates. The solid lines refer
to the fit for all events; the dashed lines correspond to the 
total background. Plot (c) shows the raw asymmetry (see text).
A requirement is applied on the event likelihood to suppress 
background.
\label{fig2}
}
\end{center}
\end{figure}

The central values of \sf and \cf were hidden until the analysis was complete.
From a data sample of 33,058 \Bztokspp candidates, we find $N_S = 117\pm 27$
signal decays with \scp = \finalscp and \ccp = \finalccp. 
The linear correlation coefficient between the two \CP parameters is 2\%.
The yield of charmless $B$ background is consistent with zero.
Figure~\ref{fig1} shows the distributions of the event variables
\mes, $\Delta E/\sigma(\Delta E)$, and $NN$, and
Fig.~\ref{fig2} shows the \deltat distributions for
the $B^0$- and the $\bar{B}^0$-tagged subsets with the raw asymmetry 
$[N_{B^0} - N_{\bar{B}^0}]/[N_{B^0} + N_{\bar{B}^0}]$.
The $N_{B^0}$ ($N_{\bar{B}^0}$) is the number of $B^0$
($\bar{B}^0$) -tagged events.
In all plots data are displayed together with the result from the fit after applying 
a requirement on the ratio of signal likelihood to signal-plus-background likelihood 
(computed without the variable plotted) to reduce the background. 

\section{SYSTEMATIC STUDIES}
\label{sec:Systematics}
We consider systematic uncertainties listed in Table~\ref{syst}.
These include the uncertainties in the parameterization of PDFs for
signal and backgrounds which were evaluated by varying parameters
within one standard deviation or using alternative shape functions.
The largest uncertainty for \cf is caused by the $NN$ shape for
continuum inside the Dalitz plot ($\sigma(\cf ) = 0.10$) and
for \sf from the 2-D parameterization ($\sigma(\sf ) = 0.04$).
We consider uncertainties in the background fractions and \CP 
asymmetry in the charmless $B$ background,
the parameterization of the \deltat resolution function and
the vertex finding method, knowledge of the event-by-event beam spot
position, imprecision in the SVT alignment, and the possible interference
between the suppressed $\bbar\to\ubar\c\dbar$ amplitude with the favored
$\b\to\c\ubar\d$ amplitude for tag-side $B$-decays~\cite{kirkby}.
We fix $\tau_{\Bz} = 1.532$~ps and $\Delta m_d = 0.505$~ps$^{-1}$
and vary them by one standard deviation~\cite{Eidelman:pdg2004}.
We correct for the small fit bias which is determined from repeated fits
to simulated events for signal and backgrounds mixed together with the
expected yields, and the uncertainty of the method is accounted for as 
systematic error.

We perform several consistency checks, including the measurement of the
\Bz lifetime; we obtain $\tau_{\Bz} = 1.25\pm 0.47$~ps. We embed different
$B$ background samples from Monte-Carlo simulation in the data sample
and obtain consistent yields and \CP parameters from the fit.
We use the PDFs to generate signal and background samples and find
that 47\% of the simulated experiments had likelihood 
values greater than the one obtained in the fit to the data.

\begin{table}[!htb]
\caption{Sources of systematic uncertainty on \sf and \cf. 
The total error is obtained by summing the individual 
errors in quadrature.}
\begin{center}
\begin{tabular}{|l|c|c|}
\hline
Source & $\sigma (\sf)$ & $\sigma (\cf)$ \\
\hline
PDF parameterization for signal and background  & 0.05  & 0.11 \\
Background fractions                            & 0.03  & 0.02 \\
\CP in charmless $B$ background                 & 0.03  & 0.01 \\
Vertex finding/Resolution function              & 0.02  & 0.05 \\
Beam spot position                              & 0.00  & 0.00 \\
SVT alignment                                   & 0.02  & 0.01 \\
Tag side interference                           & 0.00  & 0.01 \\
$\Delta m_d$, $\tau_B$                          & 0.02  & 0.01 \\
Fit Bias                                        & 0.04  & 0.02 \\
\hline
Total systematic error                          & 0.08  & 0.13 \\
\hline
\hline
\end{tabular}
\end{center}
\label{syst}
\end{table}

\section{SUMMARY}
\label{sec:Summary}
We have presented a preliminary measurement of the \CP violating 
asymmetries in \Bztokspp ($\KS\to\pi^+\pi^-$) decays
reconstructed from a sample of approximately 227 million $B\bar{B}$ pairs. 
From an unbinned extended maximum likelihood fit we obtain
$\scp = \finalscp$ and $\ccp = \finalccp$. 
The change in the log-likelihood when we fix the values of 
$-\scp$ to the average \stwob measured in $\b\to\c\cbar\s$ modes,
$\stwob =  0.725 \pm 0.037$~\cite{hfag}, and $\ccp$ to zero,
and re-fit the data sample is 2.5.
The signal yield is consistent with our findings in
the $B^0\to\KS\pi^+\pi^-$ decay~\cite{kelly} assuming the
dominant charmless final states are $f_0(980)\KS$,
$K^*(892)\piz$, $K^*_0(1430)\piz$, and non-resonant 
$\KS\piz\piz$, and isospin symmetry.

\section{ACKNOWLEDGMENTS}
\label{sec:Acknowledgments}

\input acknowledgements

\end{document}

%% file: authors_conf05013.tex
\begin{center}
\small

The \babar\ Collaboration,
\bigskip

B.~Aubert,
R.~Barate,
D.~Boutigny,
F.~Couderc,
Y.~Karyotakis,
J.~P.~Lees,
V.~Poireau,
V.~Tisserand,
A.~Zghiche
\inst{Laboratoire de Physique des Particules, F-74941 Annecy-le-Vieux, France }
E.~Grauges
\inst{IFAE, Universitat Autonoma de Barcelona, E-08193 Bellaterra, Barcelona, Spain }
% ...

\end{center}\newpage

%% file: acknowledgements.tex
We are grateful for the 
extraordinary contributions of our \pep2\ colleagues in
achieving the excellent luminosity and machine conditions
that have made this work possible.
The success of this project also relies critically on the 
expertise and dedication of the computing organizations that 
support \babar.
The collaborating institutions wish to thank 
SLAC for its support and the kind hospitality extended to them. 
This work is supported by the
US Department of Energy
and National Science Foundation, the
Natural Sciences and Engineering Research Council (Canada),
Institute of High Energy Physics (China), the
Commissariat \`a l'Energie Atomique and
Institut National de Physique Nucl\'eaire et de Physique des Particules
(France), the
Bundesministerium f\"ur Bildung und Forschung and
Deutsche Forschungsgemeinschaft
(Germany), the
Istituto Nazionale di Fisica Nucleare (Italy),
the Foundation for Fundamental Research on Matter (The Netherlands),
the Research Council of Norway, the
Ministry of Science and Technology of the Russian Federation, and the
Particle Physics and Astronomy Research Council (United Kingdom). 
Individuals have received support from 
CONACyT (Mexico),
the A. P. Sloan Foundation, 
the Research Corporation,
and the Alexander von Humboldt Foundation.

%% file: paper.bbl
\begin{thebibliography}{99}
\bibitem{cc} Unless explicitly stated, conjugate decay modes are
assumed throughout this paper.
\end{thebibliography}
